\documentclass[12pt,a4paper,onecolumn]{article}

\usepackage{amsmath}
\usepackage{amssymb}
\usepackage[utf8]{inputenc}
\usepackage{graphicx}
\usepackage{euscript}
\usepackage{longtable}
\usepackage{array}
\usepackage{topcapt}
\usepackage{caption,wrapfig}
\usepackage[usenames]{color}
\usepackage{colortbl}
\usepackage{xcolor}
\usepackage[T1]{fontenc}
\usepackage{authblk}

\title{Electron correlations and silicon nanocluster energetics}
\author[1,2]{N.L. Matsko}
\author[1,2]{Yu.A. Uspenskii}
\author[3]{E.V. Tikhonov}
\author[1,2]{V.S. Baturin}
\author[1,2]{S.V. Lepeshkin}

\affil[1]{\footnotesize P.N. Lebedev Physical Institute, Russian Academy of Sciences, Leninskii prosp. 53, 119991 Moscow, Russia}
\affil[2]{Moscow Institute of Physics and Technology - Dolgoprudny, Moscow Region 
141700, Russia}
\affil[3]{Lomonosov Moscow State University, Leninskie Gory, Moscow, 119991, Russia}

\date{}

\begin{document}
\maketitle

\begin{abstract}
The first-principle prediction of nanocluster stable structure is often hampered by the existence of many isomer configurations with energies close to the ground state. This fact attaches additional importance to many-electron effects going beyond density functional theory (DFT), because their contributions may change a subtle energy order of competitive structures. To analyze this problem, we consider, as an example, the energetics of silicon nanoclusters passivated by hydrogen Si$_{10}$H$_{2n}$ with $0\le n\le 11$, the structure of which varies with passivation from compact to loose-packed, similar to branching polymers. Our calculations performed by the DFT, hybrid functionals and Hartree-Fock (H-F) methods, as well as by the GW approximation (GWA), confirm a considerable sensitivity of structure prediction and isomer energy ordering to many-electron effects and show some results which may be obtained with the methods less computationally demanding than the GWA. 
\end{abstract}

\section{Introduction}
The unique properties of semiconductor nanoparticles are highly promising for many applications such as optoelectronics, nanoelectronics, solar cells, biosensors, etc \cite{abs1}-\cite{abs5}, so investigations on them rank among the most burning topics. One of the challenging problems is the atomic structure of nanoclusters and small nanoparticles, which generally differs a lot from the structure of bulk samples and varies widely with cluster size and composition. Nanoobject structure strongly affects its properties. There is a general understanding that variations in the atomic structure of nanoclusters are conditioned by surface atoms, that contribute to cluster stability less than central atoms. The minimum of the total energy triggers atom rearrangement being individual for each cluster. Because the experimental determination of nanocluster structure is still problematic, so the reliable first-principles structure prediction is among the hottest problems of nanocluster physics.

First-principles methods based on density functional theory received general recognition as a reasonably precise approach available for a wide class of materials. In particular, they are intensively used in the study of nanoclusters and nanomaterials, including stable structure prediction \cite{abs6}-\cite{abs8}. A relative simplicity of DFT equations and a reasonable accuracy of ground-state properties calculated allow a detailed consideration of complicated nanoobjects and nanosystems. From a mathematical point of view, the determination of cluster structure is reduced to a search for the atomic configuration realizing the global minimum of cluster energy. The search for the global minimum is especially difficult when a system has many local minima lying slightly above the global one. In this case any inaccuracy or a small systematic error can distort a subtle energy order of atomic configurations. Of course, there is a limit of accuracy, after which further improvements lose their meaning. For instance, if the energies of isomer and ground-state configurations are very close $E_{\mathrm{isomer}}-E_{\mathrm{ground}}\le k_{\mathrm{B}}T_{\mathrm{eff}}$  $\sim$ 0.03 eV - 0.04 eV ($T_{\mathrm{eff}}$  is the temperature of cluster synthesis or room temperature), both configurations have comparable chances to exist and the choice of the ground state is conventional.

In first-principles calculations, an evident source of systematic errors is the exchange-correlation (xc) contribution to the total energy. This contribution varies significantly depending on approximation employed for its description from the standard LDA and GGA approximations \cite{abs9},\cite{abs10} to the beyond-DFT methods. The important question of the first-principles structure prediction is, whether errors introduced by an approximated description of exchange and correlations shift the energies of all competitive configurations by nearly the same quantity or the shift is individual for each configuration. In the former case, the prediction of stable cluster structure is not sensitive to exchange-correlation approximations, while in the latter one a proper description of many-electron effects is of prime significance.

\begin{figure}[h!]
\centering
\includegraphics[width=0.7\textwidth]{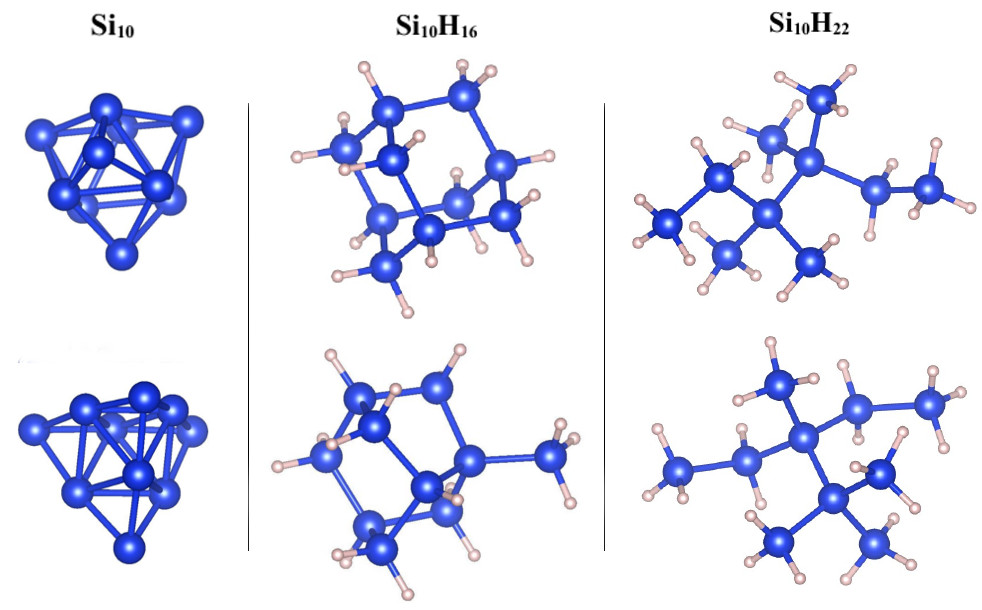}
\caption{Si$_{10}$, Si$_{10}$H$_{16}$, Si$_{10}$H$_{22}$ ground state structures (upper line) and closest isomers (bottom  line). Big dark balls - silicon atoms, small pale balls - hydrogen atoms }
\label{fig1}
\end{figure}

To elucidate this question, we make the total energy calculations of silicon nanoclusters passivated by hydrogen with formula Si$_{10}$H$_{2n}$ ($0\le n\le 11$) using different xc approximations. According to our early first-principles studies \cite{ourEPL},\cite{our2} the equilibrium structure of these clusters varies widely from very compact (Si$_{10}$) to loose-packed, similar to branching polymers built of SiH$_2$ monomers (Si$_{10}$H$_{22}$) (figure 1). Characteristic energy differences between atomic configurations also vary greatly with the hydrogen passivation. As an example, in bare Si$_{10}$ clusters the first isomer energy lies at 0.6 eV above the ground state, while in Si$_{10}$H$_{22}$ clusters this quantity falls to 0.04 eV. This diversity of structures and energetics renders the Si$_{10}$H$_{2n}$ family very suitable to study the effect of exchange-correlation refinements on cluster structure prediction. The calculations were performed for both the ground-state and low-energy isomer configurations using the GGA, hybrid functionals, Hartree-Fock and GW methods. In all, 31 cluster configurations corresponding to seven cluster compositions were calculated, that gave valuable information about the impact of many-electron effects on cluster structure prediction. It could be mentioned, that hybrid functionals  are often considered as a very accurate approach for nanoclusters description, especially B3LYP (see the works\cite{abs11}-\cite{abs15}). On the other hand these works give no reasons of the validity of such approaches.

The paper proceeds as follows. Section 2 gives the basic formulas and discusses the physical meaning of approximations relating to many-electron processes. Section 3 considers the details of computation, while Section 4 checks the precision of our GW calculations using the simplest molecules and clusters as examples. Section 5 presents the total energies of Si$_{10}$H$_{2n}$ nanoclusters in their ground-state and isomer configurations. These energies were calculated by the GGA, hybrid functional, Hartree-Fock and GW methods. The discussion of these results reveals the sensitivity of structure prediction to refinements in the description of electron exchange and correlations.

\section{Theory}

The total energy of an electron system is given by the DFT expression \cite{abs9} as:
\begin{equation} \label{t1} E_{tot}[\rho]=\left [\sum_i f_i \epsilon_i -\int \mathrm d\mathbf r\rho(\mathbf r)V_{eff}(\mathbf r)\right ] + \int \mathrm d\mathbf r\rho(\mathbf r)V_{ext}(\mathbf r) + \frac12 \int \frac {\mathrm d\mathbf r \mathrm d\mathbf r'\rho(\mathbf r)\rho(\mathbf r')e^2}{|\mathbf r-\mathbf r'|} + E_{xc}[\rho] \end{equation}
The first term in the right part of (\ref{t1}) (in square brackets) is the kinetic energy ($E_{\mathrm{kin}}$), where $\epsilon_i$ and $f_i$ are the eigenvalues of the Kohn-Sham equation and their occupation numbers. The following terms are, respectively, the energy of interaction with external field ($E_{\mathrm{ext}}$), the Hartree energy ($E_{\mathrm H}$), and the exchange-correlation energy ($E_{\mathrm{xc}}$). The effective potential of the Kohn-Sham equation is $V_{\mathrm{eff}}(\mathbf r)=V_{\mathrm{ext}}(\mathbf r)+V_{\mathrm H}(\mathbf r)+V_{\mathrm{xc}}(\mathbf r)$, where $V_{\mathrm H}(\mathbf r)=\delta E_{\mathrm H}/\delta \rho(\mathbf r)$ and $V_{\mathrm{xc}}(\mathbf r)= \delta E_{\mathrm{xc}}/\delta \rho(\mathbf r)$. The standard density functional theory describes the functional $E_{\mathrm{xc}}[\rho]$ in a local or semi-local manner. In the DFT GGA:
\begin{equation} \label{t2} E_{\mathrm{xc}}^{\mathrm{GGA}}[\rho]= \int d\mathbf r \rho(\mathbf r)\epsilon_{\mathrm{xc}} (\rho(\mathbf r),\nabla\rho(\mathbf r)) \end{equation}

whereas the DFT LDA has no dependence on density gradient. The functional (\ref{t2}) assumes that the density of exchange-correlation energy $\epsilon_{\mathrm{xc}}(\mathbf r)$ is a function of electron density $\rho(\mathbf r)$ and its gradient at the same point $\mathbf r$. This approximation is best for metals, as their screened Coulomb interaction is short-range, that makes xc interaction rather local. In semiconductors and dielectrics the screened Coulomb interaction decreases very slowly with distance, approximately as $e^2/ (\epsilon_0|\mathbf r-\mathbf r’|)$, where $\epsilon_0$ is the static dielectric constant. For this reason the use of local and semi-local approximations (the DFT LDA and GGA) is not well-justified for dielectric materials.

The Hartree-Fock approximation accounts for a long-range nature of Coulomb interaction exactly, ignoring, however, its screening. This approximation provides the exact exchange energy in terms of the density matrix $\rho(\mathbf r,\mathbf r’)$: 

\begin{equation} \label{t3} E_{\mathrm x}^{\mathrm{HF}}=-\frac12 \int \mathrm d\mathbf r \mathrm d\mathbf r' \frac{\rho(\mathbf r,\mathbf r') \rho(\mathbf r',\mathbf r) e^2}{|\mathbf r-\mathbf r'|}\end{equation}

but assumes that the contribution of electron correlations to energy is zero. This inadequacy is partially compensated by hybrid functionals, which take into account a long-range nature of interaction between electrons and describe approximately contribution from their correlations. The simplest hybrid functional which realizes this idea \cite{becke} is:
\begin{equation} \label{t4} E_{\mathrm{xc}}^{\mathrm{hyb}}=\alpha_{\mathrm{mix}} E^{\mathrm{HF}}+(1-\alpha_{\mathrm{mix}})E_{\mathrm x}^{\mathrm{GGA}}[\rho] +E_{\mathrm c}^{\mathrm{GGA}}[\rho]\end{equation}
This equation can be derived within the framework of the adiabatic connection formalism (see, for review, \cite{kummel} and references there in). Its main contributions may be treated as follows. The first term of (\ref{t4}) describes a long-range contribution to exchange energy arising from statically screened Coulomb interaction between electrons with $\epsilon_0=1/\alpha_{\mathrm{mix}}$. The second term is a short-range exchange contribution given in a semi-local manner, while the third term represents contribution from electron correlations, which also is semi-local as is required by a short-range nature of correlations. The hybrid functional (\ref{t4}) depends only on one free parameter $\alpha_{\mathrm{mix}}$, which is 0.25 (the PBE0 \cite{PBE0_1},\cite{PBE0_2} functional) or is taken around this value. In most cases the precision of (\ref{t4}) exceeds that of the LDA and GGA. More sophisticated hybrid functionals B3PW91, B3LYP, HSE \cite{B3PW91}-\cite{HSE} and others, which are common in the practical use, have three and more parameters that gives them additional flexibility and higher accuracy in the description of exchange and correlation.

The GW approximation takes into account both the exact exchange and electronic correlations, including static and dynamic ones, and disregards vertex corrections. In this approximation the energies of electronic quasiparticles $E_i$ are the eigenvalues of Dyson’s equation:
\begin{equation} \label{t5} \left[\frac{p^2}{2m}+V_{\mathrm{ext}}(\mathbf r)+V_{\mathrm H}(\mathbf r) \right]\phi_i(\mathbf r)+ \int \mathrm d\mathbf r'\Sigma_{xc} (\mathbf r,\mathbf r',E_i)\phi_i(\mathbf r') = E_i \phi_i(\mathbf r)\end{equation}
In the GWA the self-energy operator (SEO) of this equation is given by its simplest expression:
\begin{equation} \label{t6} \Sigma_{\mathrm{xc}} (\mathbf r,\mathbf r',E) = \frac{i}{2\pi}\int \mathrm dE'G(\mathbf r,\mathbf r',E+E')W(\mathbf r,\mathbf r',E')\end{equation}
Here $W(\mathbf r,\mathbf r’,E)$ and $G(\mathbf r,\mathbf r’,E)$ are, respectively, the dynamically screened Coulomb interaction and the electron Green function. In many cases it is convenient to use the spectral representations of $G(\mathbf r,\mathbf r’,E)$:
\begin{equation} \label{t7} G(\mathbf r,\mathbf r’,E) = \int_{-\infty}^{\infty} \mathrm dE' \frac{A(\mathbf r,\mathbf r’,E)}{E-E'-i\delta\cdot sgn(\mu - E')}\end{equation}
where $\mu$ is the chemical potential and $A(\mathbf r,\mathbf r’,E)$ is the electron spectral function. In the basis of Kohn-Sham’s eigenstates this function is:
\begin{equation} \label{t8} A(\mathbf r,\mathbf r’,E) = \sum_{i,i'}\psi_i(\mathbf r) A_{i,i'}(E)\psi^{\dag}_{i'}(\mathbf r')\end{equation}

It is frequently assumed (see, as example, \cite{holm_arya}) that the eigenfunctions of Dayson’s equation are nearly identical to those of the Kohn-Sham equation, so $A_{i,i’}(E)\approx A_i(E)\delta_{i,i’}$. Introducing the density of quasiparticle states $A(E)=\sum_iA_i(E)$, the total 
number of electrons is expressed as:
\begin{equation} \label{t9} N = \int_{-\infty}^{\mu} \mathrm dE A(E)\end{equation}

Being an approximation, the GWA satisfies all the conservation laws when its Green function is the solution of (\ref{t5}) \cite{baym1},\cite{baym2}. In this approximation the total energy can be calculated by two methods. One of them uses the Luttinger-Ward functional $E_{\mathrm{tot}}^{\mathrm{LW}}[G]$ \cite{lut-ward}, which is similar in structure to the functional of DFT (\ref{t1}). This functional is variational, so $E_{\mathrm{tot}}$ resulting from it is accurate even when an imperfect Green function $G$ is used. The correlation contribution to $E_{\mathrm{tot}}^{\mathrm{LW}}$ is given by the functional $\Phi_c[G]$, which is reduced in the GWA to the series of ring diagrams corresponding to the random-phase approximation (RPA). The calculation of this series is not easy for real solids, particularly for nanoclusters, which hampers the practical use of $E_{\mathrm{tot}}^{\mathrm{LW}}[G]$. 

The other method employs the Galitskii-Migdal (GM) formula \cite{mig1},\cite{mig2} which gives the total energy of electrons in terms of the electron quasiparticle spectrum. This formula is derived from the equation of motion for electrons and therefore has no analogs in DFT which is merely a static theory. The exchange-correlation contribution to energy is given by the GM formula as:
\begin{equation} \label{t10} E_{\mathrm{xc}}^{GM}=\frac12\left[\int_{-\infty}^{\mu} \mathrm dE A(E)\cdot E -\sum_i f_i \epsilon_i+\int \mathrm d\mathbf r \rho(\mathbf r)V_{\mathrm{xc}}(\mathbf r) \right]\end{equation}

The assumption that the Dyson and Kohn-Sham equations have nearly identical eigenfunctions $\phi_i(\mathbf r)\approx \psi_i(\mathbf r)$ leads to the equality of their energy contributions $E_{\mathrm{kin}}$, $E_{\mathrm{ext}}$, and $E_{\mathrm{H}}$. By this means, a higher precision of $E_{\mathrm{tot}}$ in the GWA as compared to DFT is determined by the difference between $E_{\mathrm{xc}}^{\mathrm{GM}}$ (\ref{t10}) and $E_{\mathrm{xc}}[\rho]$ (\ref{t2}).

The actual GW computation meets with two involved questions. The first one is associated with the level of the self-consistency in the solution of Dyson’s equation. In this way G$_0$W$_0$ is the simplest scheme. G$_0$ is usually picked as the DFT, Hartree-Fock or hybrid functional Green function, $W_0(\mathbf r,\mathbf r’,t)=W[G_0]$ and $\Sigma_0(\mathbf r,\mathbf r’,t)=iG_0(\mathbf r,\mathbf r’,t)W_0(\mathbf r,\mathbf r’,t)$. Quasiparticle (QP) energies goes from expression $E_{\mathrm{QP}}=E_{\mathrm{QP}}^{\mathrm{DFT}}-<\psi|V_{\mathrm{xc}}|\psi>+\Sigma(E_{\mathrm{QP}}^{\mathrm{DFT}})$. The QP spectrum found at this step is much more precise than Kohn-Sham’s one. In particular, the HOMO-LUMO gap of semiconductor nanoobjects calculated in the G$_0$W$_0$ approximation is rather close to the experimental gap, while the DFT gap is two-three times narrower \cite{hyb-Lou},\cite{tikh}. Elementary improvement can be obtained by geting QP energies as $E_{\mathrm{QP}}=E_{\mathrm{QP}}^{\mathrm{DFT}}-<\psi|V_{\mathrm{xc}}|\psi>+\Sigma(E_{\mathrm{QP}})$ and making iterations of the substitutions $E_{\mathrm{QP}}$ to $\Sigma$ for next step until convergence is achieved (self-consistency in the eigenvalues or ev-scGW). Further improvement can be obtained as follows. As the dynamically screened Coulomb interaction W$(\mathbf r,\mathbf r’,E)$ is not sensitive to the variations of G, an iterative solution of Dyson’s equation is frequently obtained with the SEO $\Sigma_G(\mathbf r,\mathbf r’,t)=iG(\mathbf r,\mathbf r’,t)W_0(\mathbf r,\mathbf r’,t)$, where only the Green’s function is iterated to the self-consistency at the fixed W$_0$ (the scGW$_0$). Fully self-consisted scGW is obtained when Dyson equation is iterated both by G and W. In principle full self-consistency eliminates errors of the start point calculations and leads to the satisfaction of conservation laws for the total energy, momentum and particles number. In works \cite{27,28,29} it was noticed that fully self-consisted GW improves the G$_0$W$_0$ total energy and ionization potentials, significantly improving the DFT and hybrid functional results. However both approximations (G$_0$W$_0$ and GW$_0$) greatly decrease the body of computation and yet provide precise quasiparticle spectra. It has been observed that sometimes the simpler G$_0$W$_0$ and GW$_0$ schemes provide even better spectra than the fully self-consistent GW calculation \cite{QP1,QP2,QP3}. This fact is explained by partial cancelation between vertex correction diagrams and the self-consistency effects \cite{onida}.  According to (10), exact quasiparticles spectra give precise total energies. For this reason the present study uses invariably the G$_0$W$_0$ approximation, while the GW$_0$ is applied only for few simple nanoobjects.

The second involved question is connected with the satellite structure in $A(E)$, which arises from dynamical interaction between electrons due to plasmon exchange. These plasmon satellites manifest themselves as the peaks of $A(E)$ positioned at multiples of the plasmon energy below each quasiparticle level $E_i$. An often-used method of satellite description is the cumulant expansion, in which the Green function for an occupied state $i$ is taken as $G_i(t)=iexp\{-iE_it+C_i(t)\}$, where $E_i$ is the quasiparticle energy and $C_i(t)$ is the cumulant. This approach precisely reproduces plasmon satellites in the experimental spectra of electron photoemission \cite{satel1},\cite{satel2}. It is noticeable that the gravity center of the spectral density $\int dE A(E)\,E$ taken over occupied states remains invariant to satellite formation \cite{hedin}. This invariance implies an upward shift of quasiparticle energies, which balances the formation of low-energy satellite structure. When G$_0$W$_0$ calculation is restricted to the quasiparticle states (ignoring satellites), this upward shift gives an illusion of decreasing cohesion \cite{godby}.

To simply circumvent this difficulty, a model spectral function can be employed. The model describes an electron at the level $i$, which interacts with a plasmon having the plasma energy E$_{\mathrm{pl}}$. The spectral function of this model can be done analytically, in terms of E$_{\mathrm{pl}}$ and the renormalization factor $Z_i$ \cite{holm_arya}. Following this research and assuming that the satellite series is infinite, the total energy correction to DFT can be given as:
\begin{equation} \label{t11} E_{\mathrm{tot}}^{\mathrm{GM}}-E_{\mathrm{tot}}^{\mathrm{DFT}} =\frac12\left[\sum_if_i(E_i-\epsilon_i)-NE_{\mathrm{pl}}|ln\overline{Z}| +\int \mathrm d\mathbf r \rho(\mathbf r)V_{\mathrm{xc}} \right]-E_{\mathrm{xc}}[\rho] \end{equation}

Here $E_i$ is the quasiparticle energy calculated by the G$_0$W$_0$, N is the total number of electrons in a nanocluster, and $ln\overline{Z} =\Sigma_i \, f_i\cdot lnZ_i/N$, where $0<Z_i\le1$. In the homogeneous electron gas two values E$_{\mathrm{pl}}$ and $|lnZ|\approx 1-Z$ have opposite trends: with the growth of $\rho$ the plasma energy increases as $\rho^{\frac12}$, while $|ln\overline{Z}|$ decreases, approaching zero at $\rho\to\infty$. Because electron density averaged over a cluster varies only slightly from one low-energy configuration to other, we expect that the second term of (\ref{t11}) (the satellite contribution) is nearly invariant to atom rearrangements and affects very weakly the structures competition. This point is examined closer in Sections 3 and 4.

It is also of interest to examine correlation between nanoclusters' polarizabilities and energetics.  Dielectric properties of the system reflect its interaction with external and internal electric field, electron screening, and affect system energy as well. According to the adiabatic connection fluctuation-dissipation theorem (ACFDT)\cite{ACFDT1,ACFDT2,ACFDT3} the correlation energy can be expressed as:
   
\begin{equation} \label{6} E_{\mathrm C}= -\int_0^1 \mathrm d\lambda\int \frac{\mathrm d\omega}{2\pi}Tr\{v[\chi^{\lambda}(i\omega)- \chi^0(i\omega)]\}  \end{equation}

where $\lambda$ is dimensionless coupling constant ($\lambda$=0 for the case of noninteraction electron system and $\lambda$=1 corresponds to the real physical system), $\chi$ is the electron response function,
$v$ denotes Coulomb interaction. It can be seen, that system correlation energy increases when $\chi^{\lambda}$, being negative, increases its absolute value. In the GWA SEO (\ref{t6}), screened Coulomb W can be rewritten as $\epsilon^{-1}(q,\omega)\times v(q) $ , where $ \epsilon^{-1}(q,\omega)$ is an inverse dielectric function, thus in GWA dielectric properties of the electron system explicitly affect computation results.
According to these simple reasons it seems reasonable to expect, that for nanocluster isomers with defined chemical formula the structures with greater $|\chi^{\lambda}(i\omega)|$ have, in general, higher total energy and thus they are energetically less favorable.
Although $\epsilon$ and $\chi$ functions depend on frequency, the static polarizability can be roughly considered as representing an approximate system dielectric response. Calculation of the static polarizability is implemented in many DFT packages and is much less time consuming than calculations on the GWA level. The consideration of the relationship between Si$_7$ and Si$_{10}$H$_{2n}$ isomers total energies and their static polarizabilities will be examined in Sections 4 and 5.

\section{Computational methods}

Our density functional calculations were performed with the DFT GGA xc functional using the Quantum Espresso (QE) \cite{qe} and VASP \cite{vasp1}-\cite{vasp4} codes. The QE calculations were made using PBE pseudopotential and a plane wave basis set having the cutoff energy of 50 Ry, while VASP’s ones were done with the Projector Augmented Waves (PAW) basis set having the cutoff energy of 500 eV with appropriate pseudopotential \cite{paw1,paw2}. Computations were performed for the supercell geometry with the vacuum layer of 13 \AA\; between nanoobject replicas (section 4 argues this layer choice). The atomic structure of considered molecules and nanoclusters was found by the QE calculation, in the process of which the positions of atoms were relaxed until resulting atomic forces became less than $10^{-4}$ Ry/\AA.  In the case of silicon clusters a nontrivial determination of cluster geometry was made with the evolutionary algorithm realized in the USPEX code \cite{uspex1,uspex2}, as has been described in our previous publications \cite{ourEPL,our2}. Both the Hartree-Fock and hybrid functional (PBE0 and B3LYP) calculations were performed using the QE code with the parameters described above.

For the GWA calculations two packages were used: BerkeleyGW \cite{bgw1}-\cite{bgw3} and VASP. In both cases for all calculations the start point were DFT eigenfunctions and eigenvalues calculated in VASP for the VASP GW$_0$ and in QE for the case of BerkeleyGW G$_0$W$_0$ and ev-scGW calculations. The VASP GW algorithm performs a direct inversion of the dielectric matrix on a frequency grid. The BerkeleyGW package can employ both the direct inversion of the dielectric matrix and the Generalized Plasmon Pole (GPP) model \cite{hyb-Lou} (GPP accelerates calculations and reduces computer memory demands). The nonuniform frequency grid for the direct inversion of the dielectric matrix in the VASP computations consisted of 50 points and of 200 points for the BerkeleyGW, the dielectric matrix was cut off at 300 eV. In the BerkeleyGW GPP dielectric matrix was cut off at 6 Ry in the momentum space.
When computing the self-energy operator of the GWA, we performed summation over all occupied and 600 unoccupied electron states.

GW$_0$ and ev-scGW require notable extra resources, therefore these schemes were applied only for small nanoobjects – the Li$_2$, N$_2$, ethyl and dimethyl ether molecules. BerkeleyGW dielectric  matrix direct inversion computations were applied only for the study of the plasmon satellites in the ethyl and dimethyl ether molecules and the Si$_7$ clusters. The GW calculations of Si$_{10}$H$_{2n}$ clusters were performed with the BerkeleyGW G$_0$W$_0$ GPP approximation.

In this paper for the nanocluster isomers energy calculations using GM formula we will neglect plasmonic modifications in the spectral function. $A(\omega)$ will be considered as a number of the quasiparticle peaks. Our calculations show, that plasmon satellites carry about 15\% of the valence spectral function weight in case of Si$_7$ isomers and about 10\% in case of dimethyl ether molecule. Thus $ln\overline{Z}$ in mentioned cases is about 0.15 and 0.1 respectively (formula \ref{t11}).
For the nanocluster isomers of a given formula, E$_{\mathrm{pl}}$ can be treated similar with high accuracy. In case of the Si$_7$ nanoclusters, shift of the satellites from QP peaks is the same for all isomers with a precision better than 2\%.  For the Si$_{10}$H$_{2n}$ isomers it means that plasmonic corrections to the relative energy should be less than 0.4 eV. This estimation is very rough and precision of our calculations for the test cases will be examined in the next section.

In our work we also perform an analysis of the polarizabilities of the studied silicon nanoclusters. Polarizability values $\alpha$ were calculated using VASP and averaged over directions. Polarizabilities are measured in the relative units, where the lowest polarizability among isomers with given formula is defined as 1.
Since total energy of a system is a value determined up to a constant shift and we were interested only in the relative energies of the nanoobjects under study, further discussion will be concerned only with energy differences. All isomer energies will be counted from the ground state structure.

\section{Calculations precision and the influence of the environment}

Table 1 presents comparison of the total energies from the experiment, DFT, hybrid functional, Hartree-Fock and GM GW calculations. First and second columns present data on Li$_2$ and N$_2$ molecules dissociation energy. Third column contains information on energy difference between two C$_2$H$_6$O molecule configurations (ethyl and dimethyl ether). By reason of the difficulties of the spin polarized BerkeleyGW computations and excessive memory requirements in the VASP GW, corresponding fields for the N$_2$ and Li$_2$ are respectively blank. 

\footnotesize

\begin{tabular}{|p{3cm}|p{3cm}|p{3cm}|p{3cm}|}
\multicolumn{4}{l}{}\\
\multicolumn{4}{l}{\textbf{Table 1.} Results for the experimental, DFT, Hartree-Fock, hybrid functional and}\\
\multicolumn{4}{l}{ GW total energy calculations. PBE, PBE0, B3LYP and H-F calculations were made}\\
\multicolumn{4}{l}{ in QE package; PAW and scGW$_0$ - in VASP; G$_0$W$_0$ and ev-scGW - in BerkeleyGW.}\\
\hline   & Li$_2$  dissociation energy, eV & N$_2$ dissociation energy, eV & ethyl - dimethyl ether energy difference, eV\\
\hline             Experiment & 1.03 & 9.8 & 0.526\\
\hline             PBE / PAW & 1.375 / 1.304 & 10.15 / 10.24 & 0.493 / 0.475\\
\hline   PBE0  & 1.87 & 9.45 & 0.529\\
\hline   B3LYP & 1.71 & 9.09 & 0.479\\
\hline   QE HF    & 2.64 & 4.88 & 0.474\\
\hline   G$_0$W$_0$ & 0.697 &  & 0.504\\
\hline   ev-scGW / scGW$_0$ &  0.72 / & \qquad/ 9.71 &  0.512 / 0.56\\
\hline
\multicolumn{4}{l}{}
\end{tabular}

\normalsize

The data in table 1 show that obtained GW energies tend to modify DFT values towards the experimental ones, in case of Li$_2$ and N$_2$ dissociation this modification is little excessive. GW energies exhibit better agreement with experiment than DFT results, they are based on. Accurate calculations of the given systems require large vacuum layer, leading to the dramatic increase of the computation cost in case of the plane wave basis set. Thus the use of high parameters was restricted, especially to VASP GW$_0$.
PBE0 and B3LYP show rather bad energy values for the systems studied (and the worst results in case of H-F). For the Li$_2$ and N$_2$ dissociation PBE0 and B3LYP give significantly inaccurate values. In case of the ethyl- dimethyl ether molecules energy difference, PBE0 gives very good result, while B3LYP error is noticeably bigger than DFT one. Thus hybrid functional in a few cases may improve results comparing to DFT, but this seems rather occasional. Summarizing the results of the table 1 we can say that the used  GM GW methods give the values closest to experimental. Comparison of the BerkeleyGW schemes shows that self-consistency in the eigenvalues improves results, but G$_0$W$_0$ results are also better than DFT and hybrid functional ones

For the DFT nanocluster and molecular calculations, size of vacuum layer needed to converge is usually referred to 7-10 angstroms (for the systems with zero electrical charge) \cite{DFTvac1},\cite{DFTvac2}. Specified value of vacuum layer makes influence of the system's copies from other supercells negligible. Because of the dynamical nature of the xc interaction, GW should be sensitive to the induced dipole-dipole interactions or dispersion interaction. This interaction could be significant at distances of about 10 \AA. Besides, in practical applications nanoclusters could be embedded in a matrix and form a periodic structure, where period value would affect system properties. It is of interest to study the dependence of the system energetics convergence (in particular within GW approach) on the vacuum layer and the supercell size.

Table 2 presents energy ordering for the first four Si$_7$ isomers inside 8.5, 10.5, 13.2, 18.5, 23.8 \AA\; cubic supercells. Isomers' structure geometries were relaxed for the particular supercell. The number at the top of each column in the table 2 denotes the actual structure (same for all supercells except small variations). For the 8.5\AA\; supercell isomers relative polarizabilities are also presented.

\footnotesize

\begin{tabular}{|p{3cm}|p{1.5cm}|p{1.5cm}|p{1.5cm}|p{1.5cm}|}
\multicolumn{5}{l}{}\\
\multicolumn{5}{l}{\textbf{Table 2.} PBE, B3LYP, PBE0, BekkeleyGW G$_0$W$_0$, VASP scGW$_0$}\\
\multicolumn{5}{l}{energies and polarizabilities in relative units for the first four Si$_{7}$}\\
\multicolumn{5}{l}{isomers in the 8.5 \AA\; - 23.8\AA\; cubic supercells.}\\
\hline   isomer number & 1 & 2 & 3 & 4  \\
\hline \multicolumn{5}{|c|}{8.5 \AA \; cubic supercell} \\
\hline   DFT, eV        &  0    &  0.17  &  0.412 & 0.416 \\
\hline   PBE0, eV       &  0    &  0.214 &  0.436 & 0.441 \\
\hline   B3LYP, eV      &  0    &  0.011 &  0.246 & 0.256 \\
\hline   G$_0$W$_0$, eV & 0.944 &  0     &  0.82  & 0.833 \\
\hline   scGW$_0$, eV   & 1.207 &  0     &  0.995 & 1.024 \\
\hline   $\alpha$       & 1.472 &  1     &  1.452 & 1.215 \\
\hline \multicolumn{5}{|c|}{10.5 \AA \; cubic supercell} \\
\hline   DFT, eV        &  0    &  0.7481&  0.9878& 0.7865\\
\hline   PBE0, eV       &  0    &  0.7491&  1.0426& 0.7815 \\
\hline   B3LYP, eV      &  0    &  0.5699&  0.7484& 0.6072\\
\hline   G$_0$W$_0$, eV &  0    &  0.4669&  0.4473& 0.5452\\
\hline \multicolumn{5}{|c|}{13.2 \AA \; cubic supercell} \\
\hline   DFT, eV        &  0    &  0.8067&  0.9898& 0.8079\\
\hline   PBE0, eV       &  0    &  0.8006&  1.0458& 0.8057\\
\hline   B3LYP, eV      &  0    &  0.6252&  0.7504& 0.6311\\
\hline   G$_0$W$_0$, eV &  0    &  0.6419&  0.6854& 0.75  \\
\hline \multicolumn{5}{|c|}{18.5 \AA \; cubic supercell} \\
\hline   DFT, eV        &  0    &  0.8091&  0.9904& 0.8093\\
\hline   PBE0, eV       &  0    &  0.8134&  1.0484& 0.8148\\
\hline   B3LYP, eV      &  0    &  0.636 &  0.7525& 0.6387\\
\hline   G$_0$W$_0$, eV &  0    &  0.7406&  0.7109& 0.7769\\
\hline \multicolumn{5}{|c|}{23.8 \AA \; cubic supercell} \\
\hline   DFT, eV        &  0    &  0.8092&  0.9904& 0.8094\\
\hline   PBE0, eV       &  0    &  0.8159&  1.0487& 0.8162\\
\hline   B3LYP, eV      &  0    &  0.6381&  0.7529& 0.6399\\
\hline   G$_0$W$_0$, eV &  0    &  0.745 &  0.7017& 0.7747\\
\hline
\multicolumn{5}{l}{}
\end{tabular}

\normalsize

In case of the 10,5 \AA\ supercell and larger, the structure with the isomer number 1 has the lowest energy for all numerical schemes. Situation dramatically changes for the 8,5 \AA \; supercell case. DFT energy sequence remains almost similar and the structure 1 is still the lowest isomer. But in the GW schemes structure 2 becomes ground state isomer, with energy much lower than other isomers. System polarizabilities analysis shows, that structure 2 in the 8,5 \AA \; supercell acquires $\alpha$ much less than other structures, that correlates with relative structure stability. Such behaviour can be associated with the increase in the interaction of the clusters in neighboring supercells. In the case of B3LYP for the 8,5 \AA \ supercell, structure 2 is also low energy isomer close to the structure 1 (ground state one). PBE0 scheme mainly represents DFT results. GPP G$_0$W$_0$  and frequency-dependent scGW$_0$ approaches show identical isomer energy ordering.
For the supercells larger than 8,5 \AA , isomers' polarizabilities have small differences and 
do not give noticeable contribution to the energy ordering of clusters.

Table 2 shows that for the GW approach the increase in the supercell size leads not only to a monotonous convergence of the relative energies of the isomers. It could be seen, that structures number 2 and 3 alternately change places on the energy scale when the supercell changes from 8.5 to 10.5 \AA, from 10.5 to 13.2 \AA, from 13.2 to 18.5 \AA. Only for the supercell 18.5 \AA\ this alternation stops and for 23.8 \AA\ supercell it is possible to say that convergence is achieved. Such behavior indicates a complex nature of the decrease in screened Coulomb interaction with distance in nanoclusters.

For the supercell more than 13.2 \AA\ (about 9 \AA\ of vacuum layer) supercell size change causes DFT energies deviations less than 0.2\% and less than 1\% for the PBE0 and B3LYP. GW relative energies show slower convergence, the convergence of energy at a level of accuracy within one percent requires an increase in vacuum layer up to 15 \AA. The results show that hybrid functionals vacuum layer to convergence is half as much again DFT, GW requires vacuum layer that is twice as large as DFT.

\section{Electron correlation effects in Si$_{10}$H$_{2n}$ and isomer energy distribution}

\begin{figure}[h!]
\centering
\includegraphics[width=0.8\textwidth]{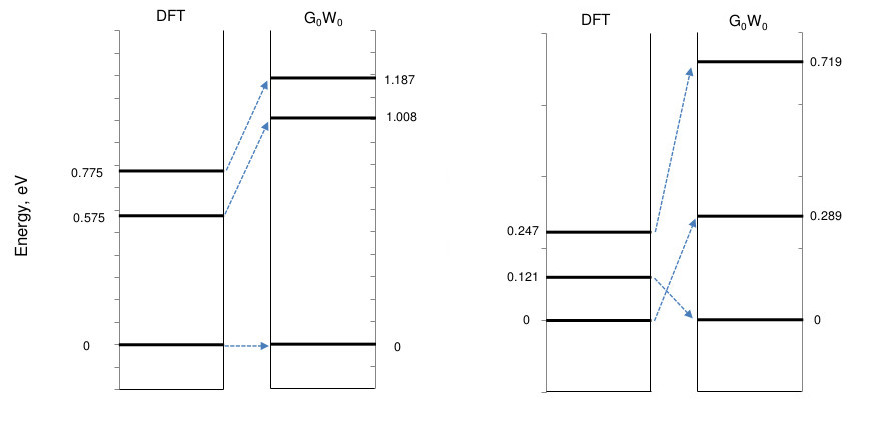}
\caption{Si$_{10}$ (left) and Si$_{10}$H$_6$ (right) clusters' isomers relative energies in DFT and G$_0$W$_0$}
\label{fig2}
\end{figure}

\begin{figure}[h!]
\centering
\includegraphics[width=0.8\textwidth]{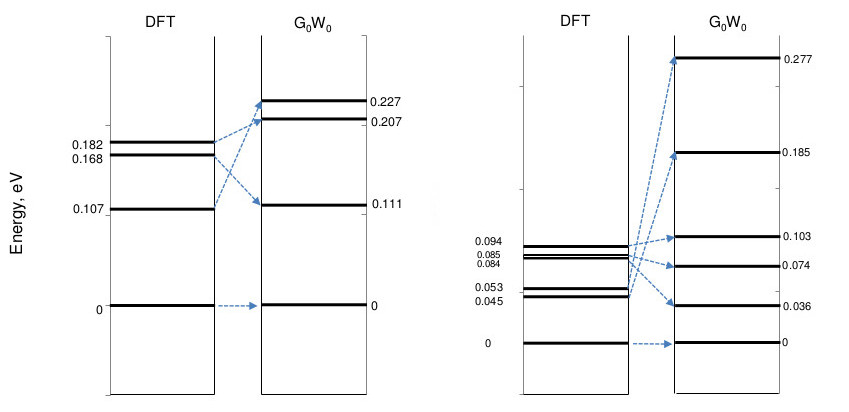}
\caption{Si$_{10}$H$_{16}$ (left) and Si$_{10}$H$_{22}$ (right) clusters' isomers relative energies in DFT and G$_0$W$_0$}
\label{fig3}
\end{figure}

\begin{figure}
\centering
\includegraphics[width=0.8\textwidth]{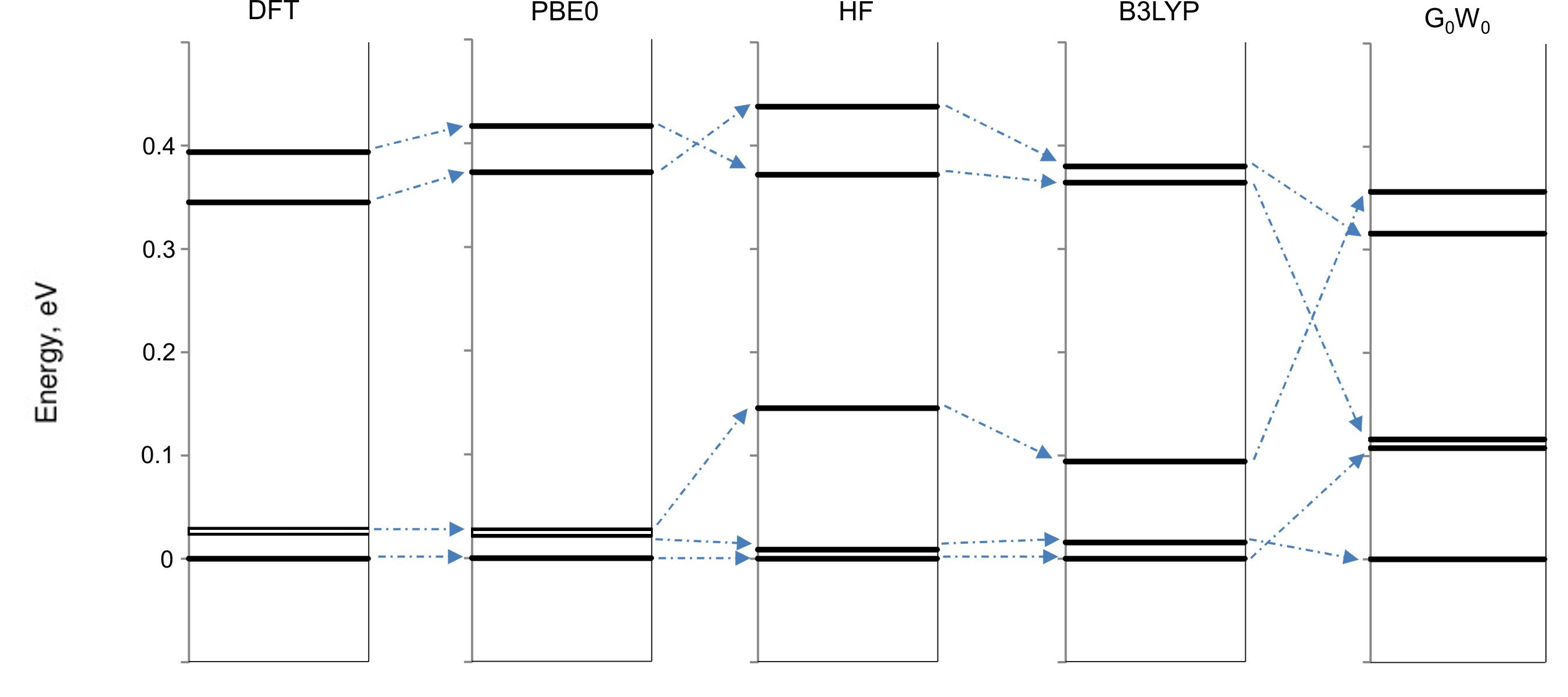}
\caption{Si$_{10}$H$_{12}$ cluster isomers relative energies in DFT, PBE0, HF, B3LYP and G$_0$W$_0$.}
\label{fig4}
\end{figure}

\begin{figure}
\centering
\includegraphics[width=0.8\textwidth]{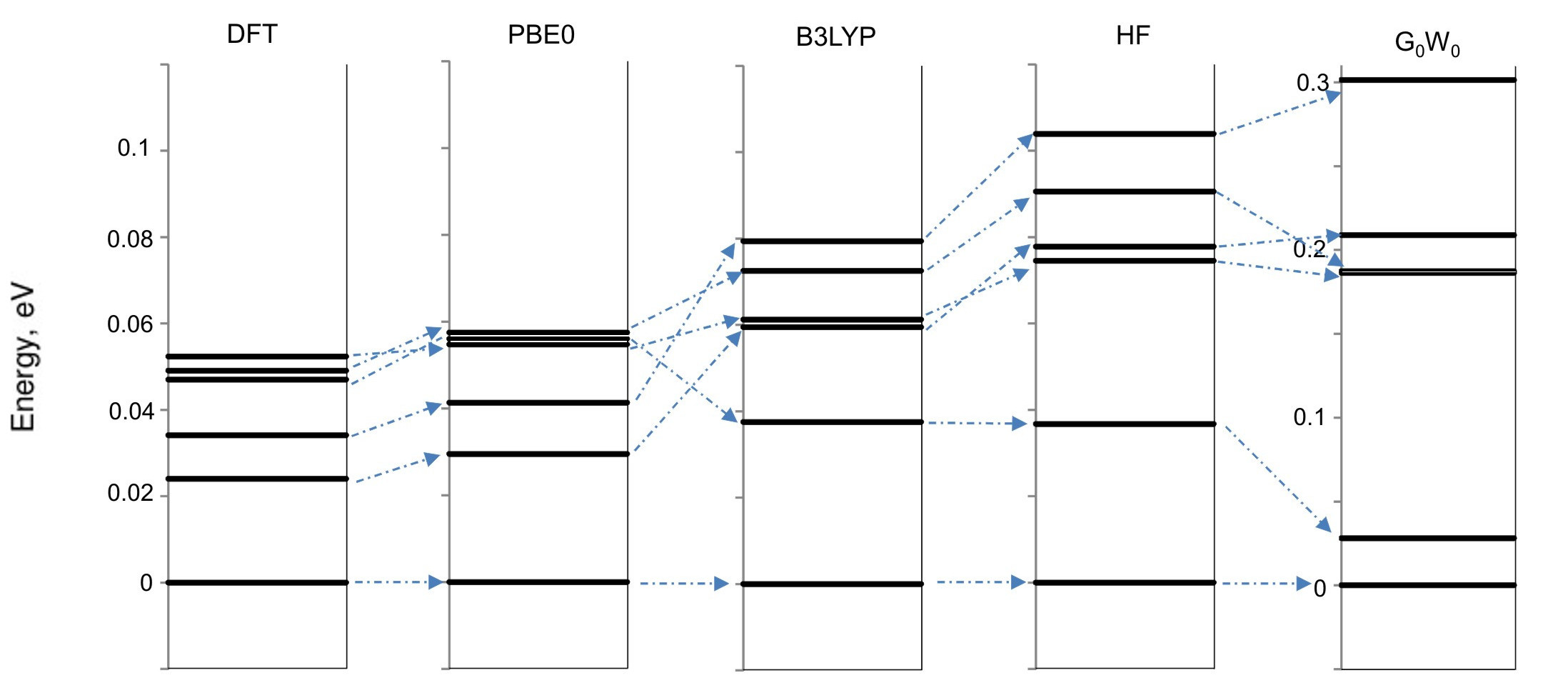}
\caption{Si$_{10}$H$_{20}$ cluster isomers relative energies in DFT, PBE0, HF, B3LYP and G$_0$W$_0$. The right graph for the G$_0$W$_0$ results has different scale.}
\label{fig4}
\end{figure}

Figures 2-5 demonstrate the relative position of energy levels for the Si$_{10}$, Si$_{10}$H$_{6}$, Si$_{10}$H$_{12}$, Si$_{10}$H$_{16}$, Si$_{10}$H$_{20}$, Si$_{10}$H$_{22}$ nanocluster isomers. The size of the vacuum layer was set to 13 angstroms. Graphs 2 and 3 present DFT PBE calculations (left parts) and BerkeleyGW G$_0$W$_0$ calculations (right parts) for the Si$_{10}$, Si$_{10}$H$_{6}$, Si$_{10}$H$_{16}$, Si$_{10}$H$_{22}$ nanoclusters. Graphs 4 and 5 present DFT PBE, PBE0, B3LYP, Hartree-Fock and G$_0$W$_0$ calculations for the Si$_{10}$H$_{12}$ and Si$_{10}$H$_{20}$ isomers.
It could be seen, that in most cases GW isomer energies change their relative ordering comparing to the DFT results. PBE0 energies mainly represent DFT ones. B3LYP and Hartree-Fock calculations exhibit some differences from DFT but also give no results consistent to the GW (even on a qualitative level). We also made PBE0 calculations with different mixing constant $\alpha_{mix}$ for the xc term (see formula \ref{t4}). Computations with $\alpha_{mix}$ varied from 0 to 1 just reproduce the results close to PBE or H-F and give no conceptually new results.
For the Si$_{10}$H$_{6}$ and Si$_{10}$H$_{12}$ isomers (figures 2 and 4) GW ground state structures also differ from DFT and hybrid functionals. In most cases the energy spread of the isomers in GWA is expanded comparing to DFT.

\begin{figure}
\centering
\includegraphics[width=0.9\textwidth]{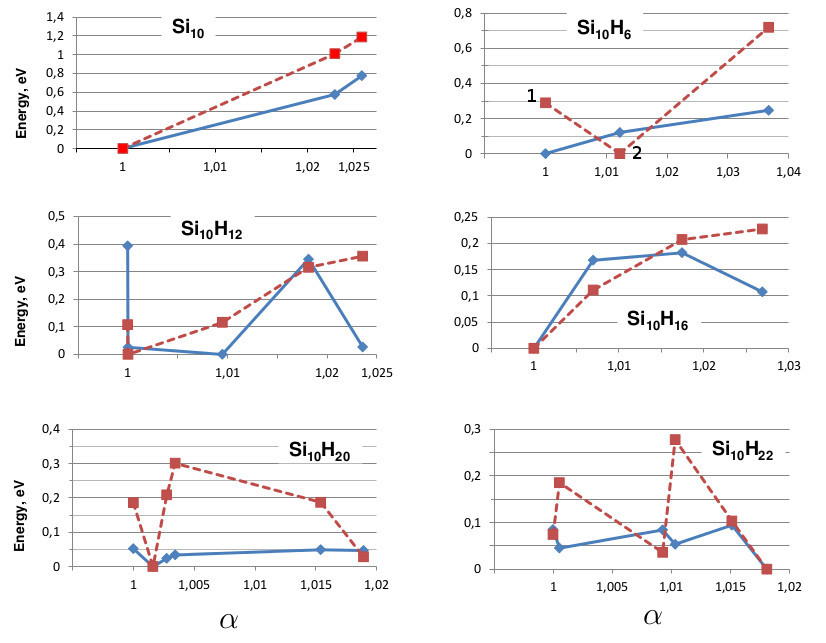}
\caption{Si$_{10}$, $Si_{10}H_6$, Si$_{10}$H$_{12}$, Si$_{10}$H$_{16}$, Si$_{10}$H$_{20}$, Si$_{10}$H$_{22}$ isomers' polarizability and total energy from DFT (rhombuses, solid line) and GM GWA (boxes, dashed line).}
\label{fig7}
\end{figure}

Figure 6 presents values of the polarizability $\alpha$ and total energy from DFT and Galitskii-Migdal G$_0$W$_0$ calculations. Each part on figure 6 presents results for the isomers with given formula: Si$_{10}$, Si$_{10}$H$_{6}$, Si$_{10}$H$_{12}$, Si$_{10}$H$_{16}$, Si$_{10}$H$_{20}$ and Si$_{10}$H$_{22}$. As can be seen, Si$_{10}$, Si$_{10}$H$_{6}$, Si$_{10}$H$_{12}$ and Si$_{10}$H$_{16}$ clusters in GM GWA mainly have lower total energy (more stable) for the isomers with lower $\alpha$. The DFT calculations do not exhibit such apparent energy-polarizability correlation. One can note violation of this rule for the Si$_{10}$H$_{6}$ isomers: for the GW energy-polarizability curve structure with $\alpha$=1 (point marked as 1) has energy of 0.3 eV higher than structure with $\alpha$=1.012 (mark 2). Calculations of the Si$_{10}$H$_{2n}$ clusters' dipole moments show, that the structure marked as 1 has dipole moment of 1.2 atomic units, the largest value of all other clusters examined. Other clusters have dipole moments of 4-100 times less. Apparently such a large dipole moment affects the energy of the structure 1, decreasing its stability.

The situation for the energy-polarizability correlation changes with the increase in the hydratation rate. For the Si$_{10}$H$_{20}$ and Si$_{10}$H$_{22}$ clusters there is no correlation between $\alpha$ and GM total energy value. This behavior can be explained as follows: for the Si$_{10}$H$_{2n}$ clusters with $n<9$ an inner Si core can be localized; increasing $n$ we get clusters of loose structure with no inner part (see figure 1). In case of Si$_{10}$H$_{20}$ and Si$_{10}$H$_{22}$ isomers the structure is branched, rather one-dimensional for each branch. In the works \cite{32,320} it was pointed out, that in small clusters microscopic dielectric properties at a few atomic distances away from the surface are almost identical to the bulk ones, whereas surface is one of the main factors that significantly change cluster polarizability. Thus Si$_{10}$H$_{2n}$ clusters with evident inner part show standard relation between system polarizability and energetics, while branched structures do not exhibit such obvious dependency.

\section{Conclusions}
DFT, hybrid functionals, Hartree-Fock, Galitskii-Migdal GW approximations were applied for the silicon-hydrogen nanoclusters' total energy computations. Precision of the methods was tested for the cases of Li$_2$, N$_2$ molecule dissociation and ethyl-dimethyl ether isomer energy difference. GM GW gives the most precise results of all methods examined, introducing a correction for DFT method, being, in turn, the starting point for the GW computation. GW energy calculations also demonstrate significantly higher sensitivity to the nanocluster environment, requiring vacuum layer to converge two times more than DFT. Moreover a non-monotonic dependence of the isomers energy distribution on supercell size was found.

Total energy calculations of the Si$_7$ and Si$_{10}$H$_{2n}$ isomers show, that correct account of the electron correlation effects is of great importance in the nanocluster systems with a big variety of structures close in energy. GWA demonstrates a notable change in the isomers' energy ordering and gives a correction to the energy of the order of tenths of eV for the isomers studied, comparing to other methods applied. Such correction will be significant for the energy ranking of the competitive structures up to a temperature of about a thousand degrees of Kelvin. It was established that standard GGA and hybrid functional methods may introduce noticeable errors into the total energy calculations for the nanoclusters consisting of tens of atoms. It is especially important for ground state structures prediction when isomers have energy differences of an order of tenths of eV or less.

It was found that, in general, the compact nanoclusters isomers with lower mean polarizability are more stable. In the branched, loose-packed structures such correlation vanishes.
This provides enough reason to expect that the minimal polarizability principle can be valid criterion for isomer stability when energy ranking large nanoclusters systems, difficult for GW calculations.

\section{Acknowledgements}

This research was supported by the Programmes of the Russian Academy of Sciences and the Russian Foundation for Basic Research (grants 16-32-00922, 16-02-00024 and 16-02-00612), the grant of the Government of the Russian Federation (14.A12.31.0003). We would like to thank A.R. Oganov for useful discussions.

\bibliography{thesis}
\bibliographystyle{gost705}

\end{document}